\documentclass[twocolumn,showpacs,aps,prl,superscriptaddress]{revtex4}

\usepackage{graphicx}
\usepackage{dcolumn}
\usepackage{epsfig}
\usepackage{amsmath}

\newcommand{\BaBarYear}    {06}
\newcommand{\BaBarNumber}  {004}

\newcommand{\SLACPubNumber} {11735}

 \newcommand{\BaBarType}      {PUB}  

\input pubboard/babarsym   

\RequirePackage{xspace}

\newcommand{\pvec}{{\bf p}}

\newcommand{\calB}{\ensuremath{{\cal B}}}

\newcommand{\DE}{\ensuremath{\Delta E}}

\newcommand{\xf}{\ensuremath{{\cal F}}}

\newcommand{\thetaT}{\ensuremath{\theta_{\rm T}}}
\newcommand{\costhr}{\ensuremath{\cos\thetaT}}

\newcommand\etal{{\it et al.}}
\newcommand{\half}{\ensuremath{{1\over2}}}

\newcommand{\bma}[1]{\boldmath{$#1$}}

\newcommand{\bfig}{\begin{figure}[htbpc!]}
\newcommand{\efig}{\end{figure}}
\newcommand\bef{\begin{figure}}
\newcommand\edf{\end{figure}}
\newcommand\dbline{\noalign{\vskip 0.10truecm\hrule}\noalign{\vskip 2pt}\noalign{\hrule\vskip 0.10truecm}}
\providecommand{\tbline}{\noalign{\vskip 0.05truecm\hrule\vskip0.05truecm}}
\newcommand\sgline{\noalign{\vskip 0.10truecm\hrule\vskip 0.10truecm}}
\newcommand\beq{\begin{equation}}
\newcommand\eeq{\end{equation}}
\newcommand\bear{\begin{array}}
\newcommand\enar{\end{array}}
\newcommand\beqa{\begin{eqnarray}}
\newcommand\eeqa{\end{eqnarray}}
\newcommand\ben{\begin{enumerate}}
\newcommand\een{\end{enumerate}}

\newcommand{\UfourS}{\ensuremath{\Upsilon(4S)}}

\newcommand{\etagg}{\ensuremath{\eta_{\gaga}}}
\newcommand{\etappp}{\ensuremath{\eta_{3\pi}}}

\newcommand{\etapepp}{\ensuremath{\etapr_{\eta\pi\pi}}}

\newcommand{\etaprg}{\ensuremath{\etapr_{\rho\gamma}}}

\newcommand{\Kst}{\ensuremath{K^*}}

   \newcommand{\rhoz}{\ensuremath{\rho^0}}

\newcommand{\fetapiz}{\ensuremath{\eta\piz}\xspace}
\newcommand{\etapiz}{\ensuremath{\Bz\ra\fetapiz}\xspace}
\newcommand{\Betapiz}{\ensuremath{\calB(\etapiz)}\xspace}

\newcommand{\Retapiz}{\ensuremath{(\retapiz)\times 10^{-6}}\xspace}
\newcommand{\uletapiz}{\ensuremath{xx}\xspace}

\newcommand{\setapiz}{\ensuremath{xx}\xspace}
   \newcommand{\fetaggpiz}{\ensuremath{\eta_{\gaga} \piz}\xspace}
   
   \newcommand{\fetappppiz}{\ensuremath{\eta_{3\pi} \piz}\xspace}

\newcommand{\fetapKz}{\ensuremath{\etapr K^0}}
\newcommand{\etapKz}{\ensuremath{\Bz\ra\fetapKz}}

\newcommand{\fetappiz}{\ensuremath{\etapr\piz}\xspace}
\newcommand{\etappiz}{\ensuremath{\Bz\ra\fetappiz}\xspace}
\newcommand{\Betappiz}{\ensuremath{\calB(\Bz\ra\etapr \piz)}\xspace}

\newcommand{\Retappiz}{\ensuremath{(\retappiz)\times 10^{-6}}\xspace}
\newcommand{\uletappiz}{\ensuremath{xx}\xspace}

\newcommand{\setappiz}{\ensuremath{xx}\xspace}
   \newcommand{\fetapepppiz}{\ensuremath{\etapr_{\eta\pi\pi} \piz}\xspace}
   \newcommand{\etapepppiz}{\ensuremath{\Bz\ra\fetapepppiz}\xspace}

\newcommand{\fetapeta}{\ensuremath{\etapr\eta}}
\newcommand{\etapeta}{\ensuremath{\Bz\ra\fetapeta}}
\newcommand{\Betapeta}{\ensuremath{\calB(\etapeta)}}

\newcommand{\Retapeta}{\ensuremath{(\retapeta)\times 10^{-6}}}
\newcommand{\uletapeta}{\ensuremath{xx}}

\newcommand{\setapeta}{\ensuremath{xx}}
  \newcommand{\fetaprgetagg}{\ensuremath{\etaprg\etagg}}
  \newcommand{\fetapeppetagg}{\ensuremath{\etapepp\etagg}}
  \newcommand{\fetaprgetappp}{\ensuremath{\etaprg\etappp}}
  \newcommand{\fetapeppetappp}{\ensuremath{\etapepp\etappp}}

  \newcommand{\etaprgetagg}{\ensuremath{\Bz\ra\fetaprgetagg}}

  \renewcommand{\Retapiz}{\ensuremath{0.6^{+0.5}_{-0.4}\pm 0.1}}	
  \renewcommand{\uletapiz}{\ensuremath{1.3}}			
  \renewcommand{\setapiz}{\ensuremath{1.3}}			

  \renewcommand{\Retappiz}{\ensuremath{0.8^{+0.8}_{-0.6}\pm 0.1}}	
  \renewcommand{\uletappiz}{\ensuremath{2.1}}			
  \renewcommand{\setappiz}{\ensuremath{1.4}}			

  \renewcommand{\Retapeta}{\ensuremath{0.2^{+0.7}_{-0.5}\pm 0.4}}	
  \renewcommand{\uletapeta}{\ensuremath{1.7}}			
  \renewcommand{\setapeta}{\ensuremath{0.4}}			

\begin{document}



\begin{flushleft}
\babar-\BaBarType-\BaBarYear/\BaBarNumber \\
SLAC-PUB-\SLACPubNumber \\
\end{flushleft}

\title{
 \large \bf\boldmath Branching Fraction Limits for \Bz\ Decays
to \fetapeta, \fetappiz\ and \fetapiz
}

%
\author{B.~Aubert}
\author{R.~Barate}
\author{D.~Boutigny}
\author{F.~Couderc}
\author{Y.~Karyotakis}
\author{J.~P.~Lees}
\author{V.~Poireau}
\author{V.~Tisserand}
\author{A.~Zghiche}
\affiliation{Laboratoire de Physique des Particules, F-74941 Annecy-le-Vieux, France }
\author{E.~Grauges}
\affiliation{Universitat de Barcelona, Fac. Fisica Dept. ECM, Avda Diagonal 647, 6a planta, E-08028 Barcelona, Spain }
% ...
\collaboration{The \babar\ Collaboration}
\noaffiliation

\date{\today}

\begin{abstract}
We describe searches for decays to two-body
charmless final states \fetapeta, \fetappiz\ and \fetapiz\ of \Bz\
mesons produced in \epem\ annihilation.  The data, collected with the
\babar\ detector at the Stanford Linear Accelerator Center, represent
232 million produced \BB\ pairs. The results for branching fractions are, in
units of $10^{-6}$ (upper limits at 90\% C.L.): 
$\Betapeta = \Retapeta\ (<\uletapeta)$,
$\Betapiz = \Retapiz\ (<\uletapiz)$, and
$\Betappiz = \Retappiz\ (<\uletappiz)$.
The first error quoted is statistical and the second systematic.
\end{abstract}


\maketitle

We present the results of searches for neutral \B\ meson decays to
\fetapeta, \fetapiz\ and \fetappiz, with a data sample expanded by about
a factor of 2.6 over the one used for our previous measurements
\cite{PRD04,PRL04}.   
In the Standard Model (SM) the processes that contribute to these decays are
described by color-suppressed tree and one-loop gluonic, electroweak or
flavor-singlet penguin amplitudes.  
For \etappiz\ and \etapiz\ the color-suppressed tree diagram is also
suppressed by approximate cancellation between the amplitudes for the
\piz\ and for the isoscalar meson to contain the spectator quark, resulting
from the mesons' isospin couplings to the quarks.
Estimates of the branching fractions for
these modes have been obtained from calculations based on QCD
factorization \cite{BBNS,BN}, perturbative QCD (for
$\Bz\ra\eta^{(\prime)}\piz$) \cite{pQCD}, soft collinear effective
theory \cite{SCET}, and flavor-SU(3) symmetry 
\cite{CGR03,CGLRS04}.  
The expectations lie in the approximate ranges $0.2$--$1.0\times10^{-6}$ for
$\Bz\ra\eta^{(\prime)}\piz$, and $0.3$--$2\times10^{-6}$ for \etapeta.

These decays are also of interest in constraining the expected value of
the time-dependent \CP-violation asymmetry parameter $S_f$ in the decay with
$f=\etapr\KS$~\cite{CGR03,GLNQ03,GRZ04}.  
The leading-order SM calculation gives the equality
$S_{\etapr\KS} = S_{J/\psi\KS}$, where the latter has been
precisely measured \cite{sin2beta}, and 
equals \stwob\ in the SM. The \CP\ asymmetry in the charmless modes
is sensitive to contributions from new physics, but also to
contamination from sub-leading SM amplitudes.
The most stringent constraint on
such contamination in $S_{{\etapr}\KS}$ comes from the measured
branching fractions of the three decay modes
studied in this paper~\cite{CGR03,GLNQ03,GRZ04}.  Recently it has also
been suggested~\cite{GZ05,G05} that
\etappiz\ and \etapiz\ can be used to constrain the
contribution from isospin-breaking effects on the value of \stwoa\ in
$B\to\pi^+\pi^-$ decays.

The results presented here are based on data collected with the \babar\
detector~\cite{BABARNIM} at the PEP-II asymmetric $e^+e^-$
collider~\cite{pep} located at the Stanford Linear Accelerator Center.
An integrated luminosity of 211 \invfb, corresponding to
$232\times 10^6$ \BB\ pairs, was recorded at the
$\Upsilon (4S)$ resonance (center-of-mass energy $\sqrt{s}=10.58\
\gev$).

Charged particles from the \epem\ interactions are detected, and their
momenta measured, by a combination of five layers of double-sided
silicon microstrip detectors and a 40-layer drift chamber,
both operating in the 1.5~T magnetic field of a superconducting
solenoid. Photons and electrons are identified with a CsI(Tl)
electromagnetic calorimeter (EMC).  Further charged particle
identification (PID) is provided by the average energy loss ($dE/dx$) in
the tracking devices and by an internally reflecting ring imaging
Cherenkov detector (DIRC) covering the central region.

\begin{table}[btp]
\begin{center}
\caption{
Selection requirements on the invariant masses of resonances and the
laboratory energies of photons from their decay.}
\label{tab:rescuts}
\begin{tabular}{lcc}
\dbline
State		& Invariant mass (MeV)			& $E(\gamma)$ (MeV)\\
\sgline						
\piz		& $120 < m(\gamma\gamma) < 150$		& $>50$		\\
\etagg		& $490 < m(\gamma\gamma) < 600$		& $>100$	\\
\etappp		& $520 < m(\pip\pim\piz) < 570$		& $>30$		\\
\etapepp	& $910 < m(\pip\pim\eta) <1000$		& $>100$	\\
\etaprg		& $910 < m(\pip\pim\gamma) <1000$	& $>200$	\\
\rhoz		& $510 < m(\pip\pim) <1000$		& ---	\\
\dbline
\end{tabular}
\vspace{-5mm}
\end{center}
\end{table}

We establish the event selection criteria with the aid of a detailed
Monte Carlo (MC) simulation of the \B\ production and decay sequences,
and of the detector response \cite{geant}.  These criteria are designed
to retain signal events with high efficiency.  Applied to the data, they
result in a sample
much larger than the expected signal, but with well characterized
backgrounds. We extract the signal yields from this sample with a
maximum likelihood (ML) fit.

The \B-daughter candidates are reconstructed through their decays
$\piz\ra\gaga$, $\eta\ra\gaga$ (\etagg), $\eta\ra\pip\pim\piz$
(\etappp), $\etapr\ra\etagg\pip\pim$ (\etapepp), and additionally for
\fetapeta\ modes, 
$\etapr\ra\rhoz\gamma$ (\etaprg), where $\rhoz\ra\pip\pim$.
Table \ref{tab:rescuts}\ lists the
requirements on the invariant mass of these particles' final states.
Secondary
charged pions in \etapr\ and $\eta$ candidates are rejected if
classified as protons, kaons, or electrons by their DIRC, $dE/dx$, and
EMC PID signatures. 

We reconstruct the \B-meson candidate by combining the
four-momenta of a pair of daughter mesons, with a vertex constraint if
the ultimate final state includes at least two charged particles.  Since
the natural widths of the $\eta$, \etapr, and \piz\ are much smaller
than the resolution, we also constrain their masses to nominal values
\cite{PDG2004}\ in the fit of the \B\ candidate.
From the kinematics of \UfourS\ decay we determine the energy-substituted mass
$\mes=\sqrt{\frac{1}{4}s-\pvec_B^2}$
and energy difference $\DE = E_B-\half\sqrt{s}$, where
$(E_B,\pvec_B)$ is the $B$-meson 4-momentum vector, and
all values are expressed in the \UfourS\ frame.
The resolution in \mes\ is $3.0\ \mev$ and in \DE\ is 24--50 MeV, depending
on the decay mode.  We
require $5.25\ \gev<\mes<5.29\ \gev$ and $|\DE|<0.3$ GeV 
($<0.2$ GeV for \fetapeta).

Backgrounds arise primarily from random combinations of particles in
continuum $\epem\ra\qqbar$ events ($q=u,d,s,c$).  We reduce these with
requirements on the angle
\thetaT\ between the thrust axis of the $B$ candidate in the \UfourS\
frame and that of the rest of the charged tracks and neutral calorimeter
clusters in the event.  The distribution is sharply
peaked near $|\costhr|=1$ for \qqbar\ jet pairs,
and nearly uniform for $B$-meson decays.  The requirement, which
optimizes the expected signal yield relative to its background-dominated
statistical error, is $|\costhr|<0.7$--$0.9$ depending on the mode.  

\begin{figure}
\includegraphics[width=0.49\linewidth]{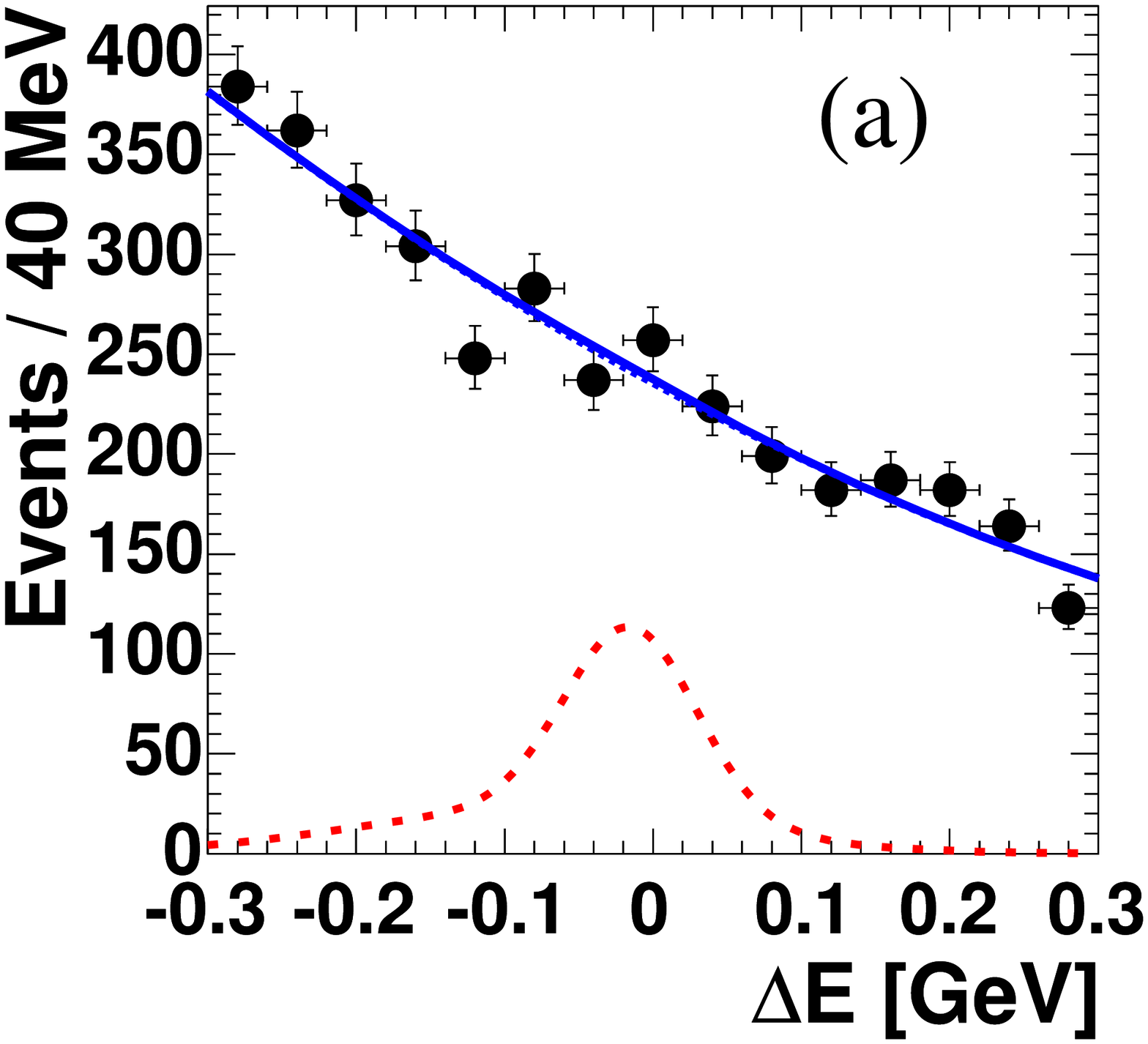}
\includegraphics[width=0.49\linewidth]{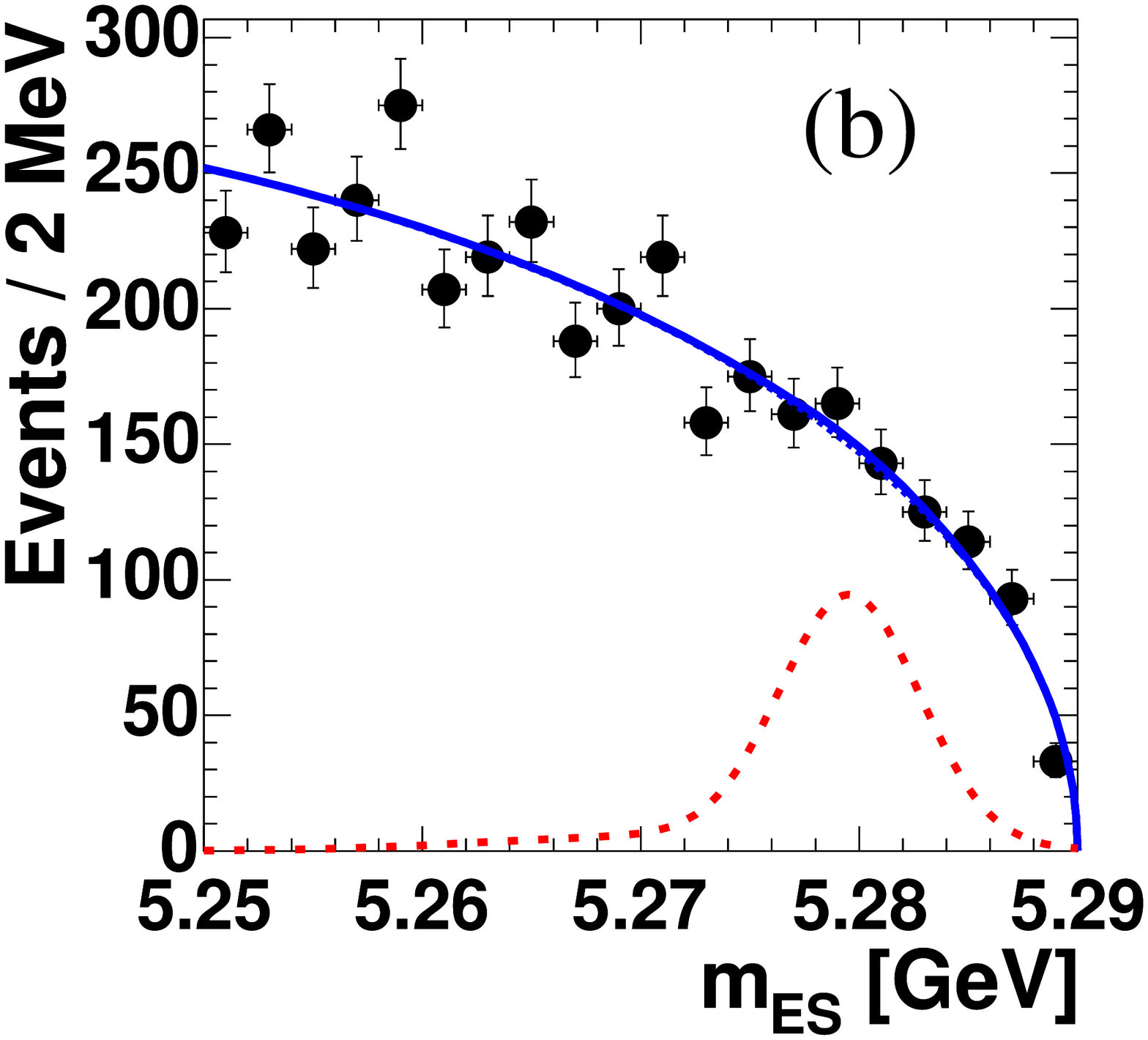}
\includegraphics[width=0.49\linewidth]{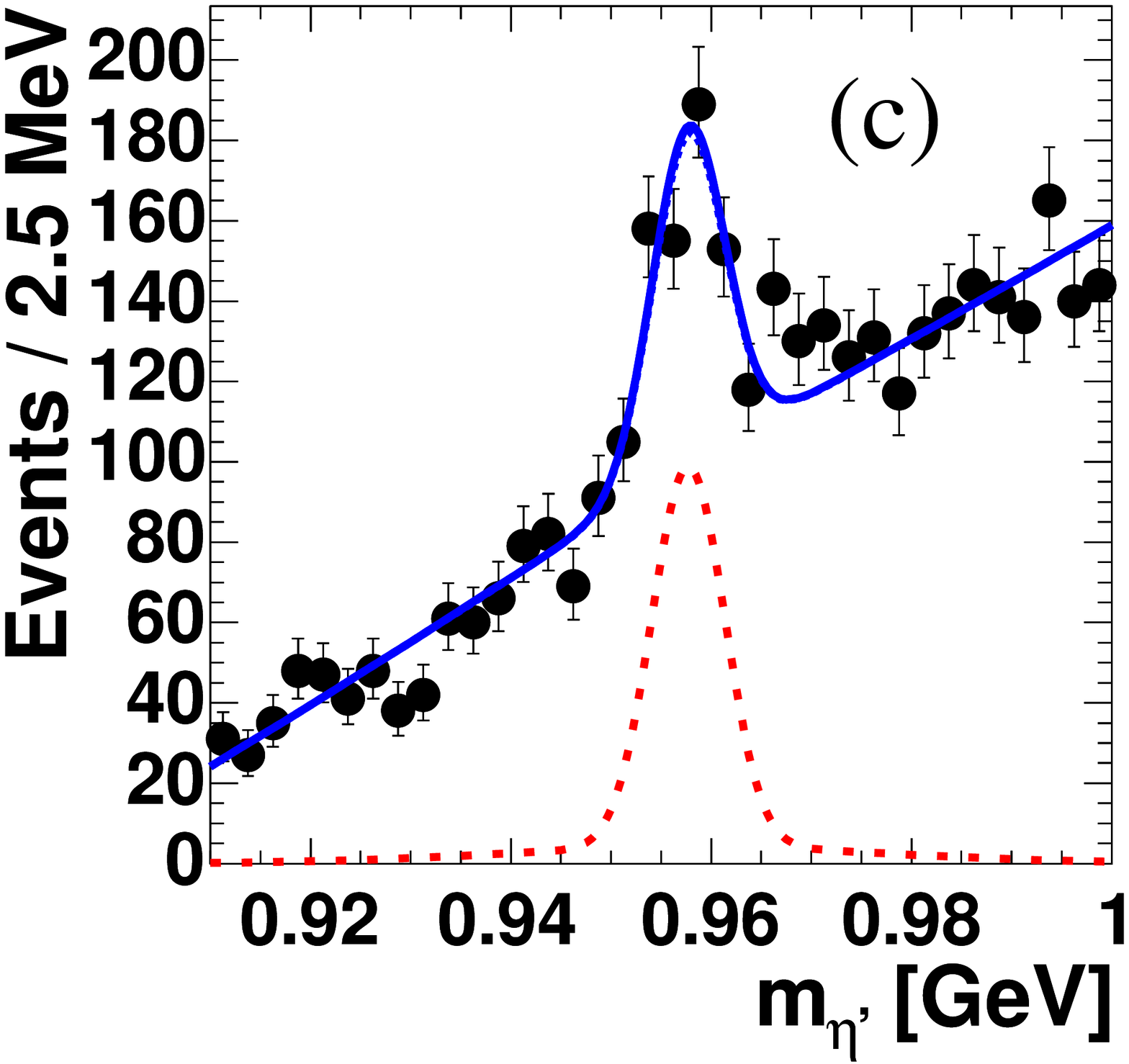}
\includegraphics[width=0.49\linewidth]{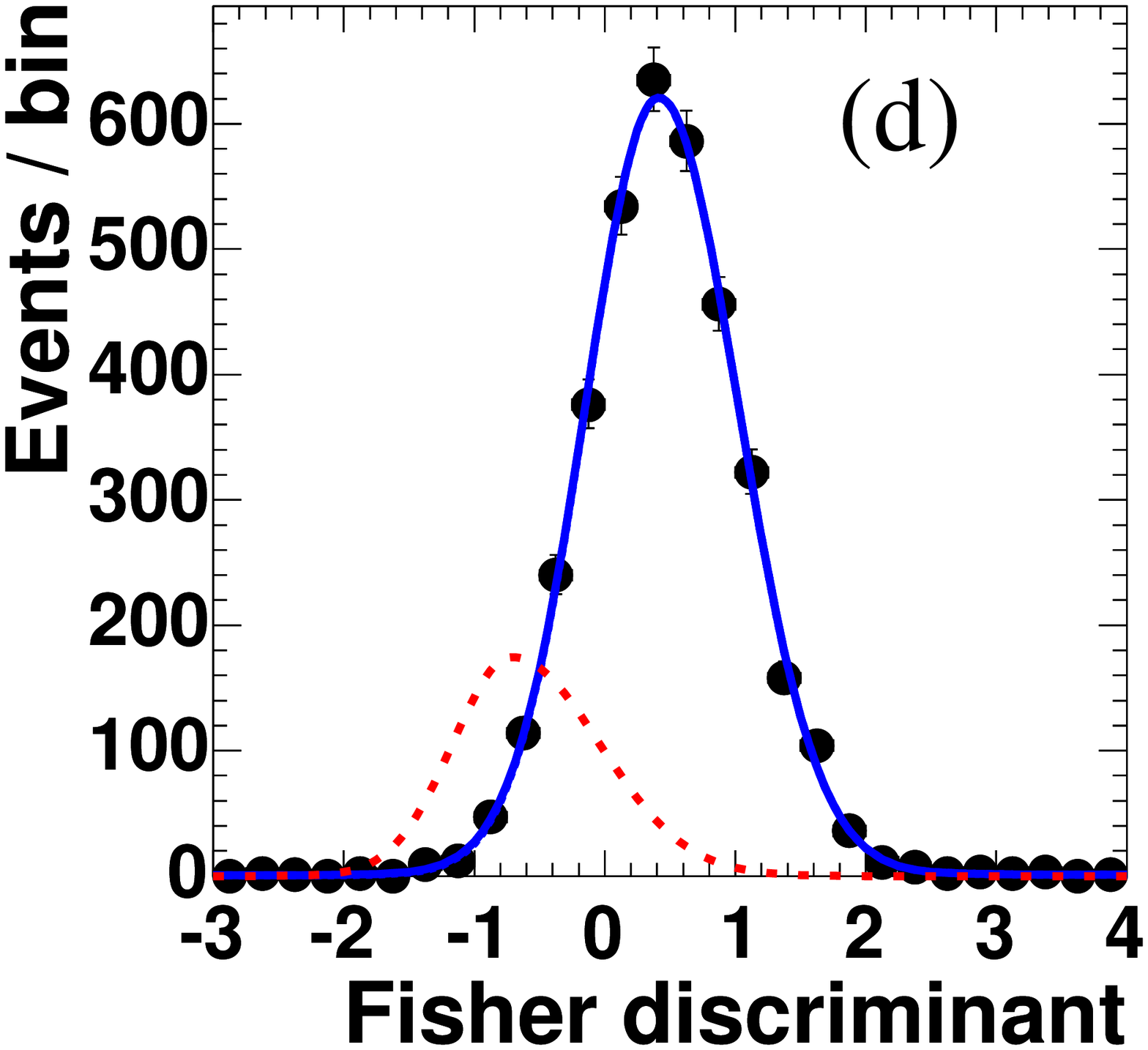}
\caption{\label{f:proj_epppiz}Plots of the \etapepppiz\ data
distribution projected on each of the fit variables:  (a) \DE, (b) \mes,
(c) $\etapr$~mass, and (d) \xf. The solid line represents the result of
the fit, 
and the dashed line the background contribution.  (The absence of signal
here nearly hides the dashed curve.)  The dotted line illustrates the
expected shape for signal, normalized arbitrarily to the data.}
\end{figure}

In the ML fit we discriminate further against \qqbar\ background with a
Fisher discriminant \xf\ that combines several variables
which characterize the energy flow in the event \cite{PRD04}.  It
provides about one 
standard deviation of separation between \B\ decay events and combinatorial
background (see Fig.~\ref{f:proj_epppiz}d).

We also impose restrictions on decay angles to exclude the most
asymmetric decays where soft-particle backgrounds concentrate and the
acceptance changes rapidly.  
We define the decay angle $\theta_{\rm dec}^k$ for a meson $k$ as the
angle between the momenta of a daughter particle and the meson's parent,
measured in the meson's rest frame.
We require for the \etaprg\
decays $|\cos\theta^{\rho}_{\rm dec}|< 0.9$ and for
$\eta^{(\prime)}\piz$ $|\cos{\theta^{\piz}_{\rm dec}}| < 0.95 $.  For
\etaprgetagg\ the requirement is $|\cos\theta^{\eta}_{\rm dec}|< 0.86$
to suppress the background $B\ra\Kst\gamma$.

The average number of candidates found per 
selected event is in the range 1.06 to 1.23, depending on the final
state.  We choose the candidate with the smallest value of a $\chi^2$
constructed from the deviations from expected values of one or more of
the daughter resonance masses.  From the simulation we find that
this algorithm selects the correct-combination candidate in about
two thirds of the events containing multiple candidates, and that it
induces negligible bias.

We obtain yields for each channel from a maximum likelihood
fit with the input observables \DE, \mes, \xf, and
$m_{1,(2)}$, the daughter invariant mass spectrum of the $\eta$ and/or
\etapr\ candidate.  
The selected sample sizes are given in the second column of
Table~\ref{tab:results}.  Besides any signal events they contain \qqbar\
(dominant) and \BB\ with $b\ra c$ combinatorial background, and a
fraction that we estimate from the simulation to be less than 0.2\%\ of
feed-across from other charmless \BB\ modes.  The latter events have ultimate
final states different from the signal, but with similar kinematics so
that broad peaks near those of the signal appear in some observables,
requiring a separate component in the probability density function
(PDF).
The likelihood function is
\begin{eqnarray}
{\cal L} &=& \exp{(-\sum_{j} Y_{j})}
\prod_i^{N}\sum_{j} Y_{j} \times \label{eq:likelihood}\\
&&{\cal P}_j (\mes^i) {\cal P}_j (\DE^i) { \cal
P}_j(\xf^i) {\cal P}_j (m_{1}^i)\left[{\cal P}_j(m_{2}^i)\right], \nonumber
\end{eqnarray}
where $N$ is the number of events in the sample, and for each component
$j$,
$Y_{j}$ is the yield of events and ${\cal P}_j(x^i)$ the
PDF for observable $x$ in event $i$. For 
the modes $\Bz\ra\etapepp\eta$ we found no need for the \BB\ background
component.   The factored form of the PDF indicated in Eq.\
\ref{eq:likelihood}\ is a good approximation, particularly for the
combinatorial \qqbar\ component, since correlations among observables
measured in the data (dominantly this component) are small.  Distortions
of the fit results caused 
by this approximation are measured in simulation and
included in the bias corrections and systematic errors discussed below.

We determine the PDFs for the signal and \BB\ background components
from fits to MC data.  We calibrate the resolutions in \DE\ and
\mes\ with large control samples of $B$ decays to charmed final states
of similar topology (e.g.\ $B\ra D(K\pi\pi)\pi$).  For the combinatorial
background the PDFs are determined in the fits to the data.  However the
functional forms are first deduced from fits of that component alone to
sidebands in (\mes,\,\DE), so that we can validate the fit before
applying it to data containing the signal.

We use the following functional forms for the PDFs: sum of
two Gaussians for ${\cal P}_{\rm sig}(\mes)$, ${\cal P}_{{\rm sig},\BB}(\DE)$,
and the sharper structures in ${\cal P}_{\BB}(\mes)$ and ${\cal
P}_j(m_k)$; linear or quadratic dependences for 
combinatorial components of ${\cal P}_{\BB,\qqbar}(m_k)$ and for ${\cal
P}_{\qqbar}(\DE)$; and a conjunction of two  
Gaussian segments below and above the peak with different widths, plus a broad
Gaussian, for ${ \cal P}_j(\xf)$.  The \qqbar\ 
background in \mes\ is described by the function
$x\sqrt{1-x^2}\exp{\left[-\xi(1-x^2)\right]}$, with
$x\equiv2\mes/\sqrt{s}$ and parameter $\xi$.  These are discussed in
more detail in \cite{PRD04}, and some of them are illustrated in
Fig.~\ref{f:proj_epppiz}.  

We allow the parameters most important for the determination of the
combinatorial background PDFs to vary in the fit, along with the yields
for all components.  Specifically, the free background parameters are
most or all of the following, depending on the decay mode: $\xi$ for
\mes, linear and quadratic coefficients for \DE, area and slope of the
combinatorial component for $m_k$, and the mean, width, and width
difference parameters for \xf.  Results for the yields are presented in
the third column of Table\ \ref{tab:results}\ for each sample.

\begin{table*}[btp]
\caption{
Number of events $N$ in the sample, fitted signal yield $Y_S$ in events (ev.), measured bias, detection
efficiency $\epsilon$, daughter branching fraction product ($\prod\calB_i$),
and measured branching fraction \calB\ with statistical error for each
decay chain, and for the combined measurements the significance~$\cal S$
(with systematic uncertainties included), branching fraction with
statistical and systematic error, and in parentheses the 90\%
C.L. upper limits.  The number of produced \BB\ pairs is
$(231.8\pm2.6)\times 10^6$. 
}
\label{tab:results}
\newcommand{\mn}{\ensuremath{\phantom{-}}}
\newcommand{\eff}{$\epsilon$ (\%)}
\newcommand{\pbf}{$\prod\calB_i$ (\%)}
\newcommand{\signf}{$\cal S$ ($\sigma$)}
\begin{tabular}{lcr@{.}lccr@{.}lcl}
\dbline
Mode	      		& $N$ (ev.)
				&\multicolumn{2}{c}{$Y_S$ (ev.)}
							& Bias (ev.)	&\eff	&\multicolumn{2}{c}{\pbf}
											&\signf	&\multicolumn{1}{c}{\calB\ $(10^{-6})$}	\\
\tbline
~~\fetapeppetappp	& 539	& $2$&$0^{+3.1}_{-2.0}$	& $1.9\pm1.0$	& 13.8	&~~~~3&95	& 	& $\mn0.1^{+2.4}_{-1.6}$ 	\\
~~\fetapeppetagg	& 1448	& $2$&$1^{+3.5}_{-2.2}$	& $0.7\pm0.4$	& 22.3	&  6&89	& 	& $\mn0.4^{+1.0}_{-0.6}$ 	\\
~~\fetaprgetappp	& 8268	& $-8$&$6^{+8.7}_{-7.0}$& $0.0\pm0.4$	& 14.9	&  6&67	&	& $-3.8^{+3.8}_{-3.0}$ 	\\
~~\fetaprgetagg		& 16861	& $1$&$5^{+10.5}_{-8.5}$& $0.0\pm0.5$	& 21.8	& 11&63	& 	& $\mn0.2^{+1.8}_{-1.4}$ 	\\
\bma{\fetapeta}		& 	&\multicolumn{2}{c}{}	& 		&	&\multicolumn{2}{c}{}& {\boldmath \setapeta}	&\mn{\boldmath \Retapeta} \quad({\boldmath $<$\uletapeta})	\\ 
~~\fetappppiz		& 2334	& $10$&$3^{+8.6}_{-6.7}$& $1.2\pm0.7$	& 16.3	& 22&6	& 	& $\mn1.1^{+1.0}_{-0.8}$ 	\\
~~\fetaggpiz		& 5493	& $6$&$5^{+11.5}_{-9.6}$& $1.2\pm0.8$	& 20.7	& 39&4	& 	& $\mn0.3^{+0.6}_{-0.5}$ 	\\
\bma{\fetapiz}		& 	&\multicolumn{2}{c}{}	& 		&	&\multicolumn{2}{c}{}& {\boldmath \setapiz}	& \mn{\boldmath \Retapiz} \quad({\boldmath $<$\uletapiz})	\\
\bma{\fetappiz}		& 3663	& $7$&$9^{+6.9}_{-5.2}$	& $1.2\pm0.6$	& 17.5	& 17&5	& {\boldmath \setappiz}	& \mn{\boldmath \Retappiz} \quad({\boldmath $<$\uletappiz})	\\
\dbline
\end{tabular}
\end{table*}

We test and calibrate the fitting procedure by applying it to
ensembles of simulated \qqbar\ experiments drawn from the PDF into which
we have embedded the expected number of signal and \BB\ background
events randomly extracted from the fully simulated MC samples. We find
biases of 0--2 events, somewhat dependent on the signal size.  The bias
values obtained for simulations that reproduce the yields found in the
data are given in the fourth column of Table~\ref{tab:results}.

\begin{figure}[bhtp]
\begin{center} 
\includegraphics[width=1.\linewidth]{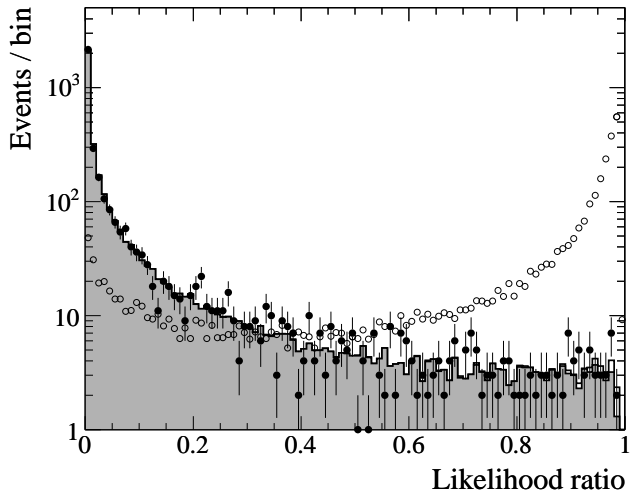}
\end{center}
\caption{\label{fig:likelihood_etapeta}%
The likelihood ratio $\calL_{sig}/[\calL_{sig}+\sum \calL_{bkg}]$ 
for \etapepppiz. The open circles represent an arbitrarily
large simulated signal component, the solid points represent the data, 
the solid histograms are from toy samples of background
(shaded) and background plus signal (white, barely visible in the
rightmost bins, given the small signal yield).  }
\end{figure}

In Fig.\ \ref{f:proj_epppiz}\ we show, as representative of the several
fits, the projections of the PDF and data for the \etapepppiz\
sample.
The goodness-of-fit is further demonstrated by the distribution of the
likelihood ratio $\calL_{sig}/[\calL_{sig}+\sum \calL_{bkg}]$ for data
and for simulation generated from the PDF model, shown for 
the same decay mode in Fig.\ \ref{fig:likelihood_etapeta}. We
see good agreement between the model and the data. By construction the
background is concentrated near zero, while any signal would appear
in a peak near one. 

We determine the reconstruction efficiencies, given in Table
\ref{tab:results}, as the ratio of reconstructed and accepted events in
simulation to the number generated.
We compute the branching fraction for each channel by
subtracting the fit bias from the measured yield, and dividing the
result by the efficiency and the number of produced \BB\ pairs~\cite{PRD04}.
We assume equal decay rates of the \UfourS\ to \BpBm\ and \BzBzb .
Table \ref{tab:results} gives the numbers pertinent to these computations.
The statistical error on the signal yield or branching fraction is taken
as the change in the central value when the quantity $-2\ln{\cal L}$
increases by one unit from its minimum value.

We combine results where we have multiple
decay channels by adding the functions
$-2\ln{\left\{\left[\calL(\calB)/\calL(\calB_0)\right]\otimes
G(\calB;0,\sigma^\prime)\right\}}$, where $\calB_0$ is 
the central value from the fit, $\sigma^\prime$ is the systematic
uncertainty, and $\otimes G$ denotes convolution with a Gaussian function.
We give the resulting final branching fractions for each mode in 
Table \ref{tab:results} with the significance, taken
as the square root of the difference between the value of $-2\ln{\cal
L}$ (with additive systematic uncertainties included) for zero signal and the
value at its minimum.  The 90\%\ C.L. upper limits are taken to be the
branching fraction below which lies 90\% of the total of the likelihood
integral in the positive branching fraction region.

The systematic uncertainties on the branching fractions arising
from lack of knowledge of the PDFs have been included in part in the
statistical 
error since most background parameters are free in the fit.  For the
signal, the uncertainties in PDF parameters are estimated from the
consistency of fits to MC and data in control modes.  Varying the
signal-PDF parameters within these errors, we estimate yield
uncertainties of 0--2 events, depending on the mode.  The uncertainty
from fit bias (Table \ref{tab:results}) includes its statistical
uncertainty from the simulated experiments, and half of the correction
itself, added in quadrature. Similarly we estimate the
uncertainty from modeling the \BB\ backgrounds by taking half of the
contribution of that component to the fitted signal yield, 0.2--1.2
events.  These additive systematic errors are dominant for these modes
with little or no signal yield.

Uncertainties in our knowledge of the efficiency, found from auxiliary
studies, include $0.8\%\times N_t$ and $1.5\%\times N_\gamma$, where
$N_t$ and $N_\gamma$ are the number of tracks and photons, respectively,
in the $B$ candidate.  The uncertainty in the total number of \BB\ pairs in the
data sample is 1.1\%.  Published data \cite{PDG2004}\ provide the
uncertainties in the $B$-daughter product branching fractions (0.7--3.9\%).
The uncertainties in the efficiency from the event selection are about~1\% .

After combining the
measurements we obtain the central values and
90\%\ C.L. upper limits for the branching fractions:
\begin{center}
\begin{tabular}{l}
$\Betapeta = (\Retapeta ) \times 10^{-6}\ (<\uletapeta\times 10^{-6})$,\\
$\Betapiz = (\Retapiz ) \times 10^{-6}\ (<\uletapiz \times 10^{-6})$, \\
and \\
$\Betappiz = (\Retappiz ) \times 10^{-6}\ (<\uletappiz\times 10^{-6})$. \\
\end{tabular}
\end{center}
We find no evidence for these decays, and our upper limits represent
two to three-fold improvement over the previous measurements
\cite{PRD04,PRL04,belle_etapi0}. 
The range of sensitivity of these measurements is comparable to the
range of the theoretical estimates.  

These results can be used to constrain the expected value of the \CP\
asymmetry $S_f$ in relation to
\stwob\ for the decay \etapKz~\cite{CGR03,GLNQ03,GRZ04}.
Using the method proposed by Gronau \etal~\cite{GRZ04}, we estimate that
our results will provide approximately $20\%$ improvement of the
prediction for the contribution of the color suppressed tree amplitude in
\etapKz\ decays.  This translates into a 20\% reduction of this
theoretical uncertainty in $S_{\etapr\KS}$.  We find a similar
improvement in the corresponding uncertainty of \stwoa\ measured with
$B\to\pi^+\pi^-$ decays~\cite{GZ05,G05}. 

We are grateful for the excellent luminosity and machine conditions
provided by our \pep2\ colleagues, 
and for the substantial dedicated effort from
the computing organizations that support \babar.
The collaborating institutions wish to thank 
SLAC for its support and kind hospitality. 
This work is supported by
DOE
and NSF (USA),
NSERC (Canada),
IHEP (China),
CEA and
CNRS-IN2P3
(France),
BMBF and DFG
(Germany),
INFN (Italy),
FOM (The Netherlands),
NFR (Norway),
MIST (Russia), and
PPARC (United Kingdom). 
Individuals have received support from CONACyT (Mexico), 
Marie Curie EIF (European Union),
the A.~P.~Sloan Foundation, 
the Research Corporation,
and the Alexander von Humboldt Foundation.

%

\renewcommand{\baselinestretch}{1}

\end{document}